\documentclass[prb,twocolumn,showpacs,preprintnumbers,amsmath,amssymb]{revtex4}

\usepackage{graphicx}
\def\ub{{\overline{u}}}
\def\vb{{\overline{v}}}

\begin{document}

\title{Critical behavior of frustrated spin systems with nonplanar orderings}

\author{Pietro Parruccini}
 \email{parrucci@df.unipi.it}
\affiliation{Dipartimento di Fisica dell' Universit\`a di 
Pisa and INFN,\\
 Via Buonarroti 2, I-56100 Pisa, Italy.}

\date{\today}

\begin{abstract}
The critical behavior of frustrated spin systems with nonplanar orderings is 
analyzed by a six-loop study in fixed dimension of an effective 
 O$(N) \times $O$(M)$ Landau-Ginzburg-Wilson Hamiltonian. For this
purpose the large-order behavior of the field theoretical expansion is 
determined.
No stable fixed point is found in the physically interesting case
 of $M=N=3$, suggesting a first-order transition in this system.
The large $N$ behavior is analyzed for $M=3, 4, 5$ and the line $N_c(d=3,M)$ which limits the region of second-order phase transition is computed.
\end{abstract}


\maketitle
%%%%%%%%%%%%%%%%%%%%%%%%inizio%%%%%%%%%%%%%%%%%%%%%%%%%%%%%%%%%%
\section{INTRODUCTION}
The critical behavior of frustrated spin systems with noncollinear order
represents an important and, at the same time, enchanting issue.~\cite{Kawamura-98,Pelissetto:2000ek,cp-97}
The debated topic concerns the nature of the phase transition which these systems should undergo. In the case of a second-order phase transition, the universality class is expected to be novel \cite{Kawamura-88,Kawamura-98,Ka-90} and characterized by new critical  exponents different from those of O($N$)-symmetric models.

Typical systems which have been studied for many years and where the situation
is still unclear are the staked triangular antiferromagnets and helimagnets,
 which present a noncollinear but planar spin orderings of the ground state.~\cite{Kawamura-88,Kawamura-98}
On this issue, there is much debate since field theoretical methods, Monte 
Carlo simulations and experiments provide contradictory results in many cases,
sometimes supporting a weak first-order phase transition, and sometimes a 
second-order one (see Refs. \onlinecite{Kawamura-98,Pelissetto:2000ek,cp-97} 
for reviews on this topic).
The perturbative field theoretical studies on canted magnetic systems are 
based on an effective O$(N) \times $O$(M)$ Landau-Ginzburg-Wilson (LGW) 
Hamiltonian with $M \leq N $, which describes $M$-dimensional spin orderings in isotropic
 $N$-spin space~\cite{Kawamura-98,Pelissetto:2000ek,Kawamura-88} with the
O$(N)/$O$(N-M)$ pattern of symmetry breaking:
\begin{widetext}
\begin{eqnarray}
{\cal H} = \int d^d x  \left\{ {1\over2}
      \sum_{a} \left[ (\partial_\mu \phi_{a})^2 + r \phi_{a}^2 \right]
+ {1\over 4!}u_0 \left( \sum_a \phi_a^2\right)^2 %\right.  \nonumber\\
% \left.
 + {1\over 4!}  v_0
\sum_{a,b} \left[ ( \phi_a \cdot \phi_b)^2 - \phi_a^2\phi_b^2\right]
             \right\},
\label{LGWH}
\end{eqnarray}
\end{widetext}
where $\phi_a$ ($1\leq a\leq M$) are $M$ sets of $N$-component vectors.
Negative values of $v_0$ correspond to simple ferromagnetic or 
antiferromagnetic ordering, and to magnets with sinusoidal
spin structures.~\cite{Kawamura-98} The 
condition $0< v_0< M/(M-1)u_0$ is required to have
noncollinearity and the boundedness of the free energy. For $M=2$ the
Hamiltonian (\ref{LGWH}) describes magnets with noncollinear but planar
orderings as the $XY$ ($N=2$) and Heisenberg ($N=3$) frustrated 
antiferromagnets, while for $M \geq 3$ systems with noncoplanar ground states.~\cite{Ka-90,Kawamura-98}

 The physical relevant case $M=N=3$ represents the noncoplanar spin orderings
which are three-dimensional in spin space and can be realized in structures 
in which the instability occurs simultaneously at three inequivalent
points in the wavevector space as shown in Fig. 1.~\cite{Kawamura-98,Ka-90}

  \begin{figure}[tb]
\includegraphics[height=3truecm,width=6truecm]{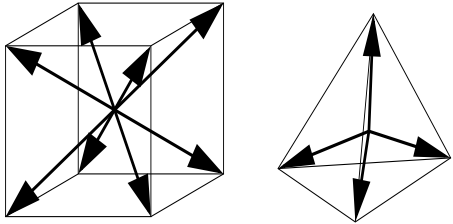}
\caption{Examples of nonplanar spin orderings ($M=3$) in fcc antiferromagnets
  of type-II (left) and type-I (right). }
\label{fcc}
\end{figure}

For $M=2$, the critical properties of the Hamiltonian (\ref{LGWH}) have been 
extensively studied in the framework of the $ \epsilon =4-d$,~\cite{Kawamura-98,asv-95,Pelissetto:2001fi} $1/N$,~\cite{Kawamura-98,Pelissetto:2001fi,Gracey:2002pm,Gracey:2002ze} $\tilde{\epsilon}=d-2$ expansions,~\cite{az-de-de-jo} as well as in fixed dimension
 $d=2$,~\cite{Calabrese:2001be,Calabrese:2002bj,Calabrese:2002af} and $d=3$,~\cite{as-94,Calabrese:2002af,Calabrese:2003ib,Pelissetto:2000ne,Pelissetto:2001hm} and with the exact renormalization group (ERG) technique.~\cite{Tissier:2001uk,tmd-99,zu-94}
The existence and stability of the fixed points were 
found to depend on $N$ in three dimensions. For
sufficiently large values of components of the order parameter four fixed
points exist in the region $u,v>0$: Gaussian, O$(2N)$, chiral and antichiral.
The chiral fixed point is the only stable one. Decreasing the value of $N$
the $\epsilon$  expansion predicts the existence of an $N_c(M=2)$ 
greater than three at which the chiral and antichiral fixed points coalesce,
 disappearing for $N<N_c(M=2)$.  These results, which are in agreement 
with the $1/N$,~\cite{Kawamura-98,Pelissetto:2001fi} and the ERG ones,~\cite{Tissier:2001uk} suggest the existence of a first-order phase
transition for $XY$ and Heisenberg frustrated antiferromagnets.
 However one should consider that both the $\epsilon$ and $1/N$ results
 are essentially adiabatic moving from  large $N$ and small $\epsilon$.
For this reason they do not necessarily
provide the (essentially nonperturbative) features of the models in the
region below $N_c(M=2)$. In fact six-loop calculations
performed directly in three dimensions~\cite{Pelissetto:2000ne,Calabrese:2003ib} showed a different scenario characterized by two critical numbers of 
components of the order parameter $N_c(M=2)=6.4(4)$ and $N_{c2}(M=2)=5.7(3)$:
 for $N_{c2}(M=2)<N<N_c(M=2)$ no stable fixed point is present in the RG flow,
 while for $N<N_{c2}(M=2)$ a stable  chiral fixed point appears again,~\cite{Pelissetto:2000ne} which is
 a focus~\cite{Calabrese:2002af} not related with its counterpart found within
 the $\epsilon$ and $1/N$ expansions.
The discrepancy between the ERG predictions and the weak-coupling ones remains
an open problem which could be
explained either with the low order considered in the derivative expansion 
in the first case, or by the problem of Borel-summability in the second one.

For generic values of $M\geq 3$, the $\epsilon$ and $1/N$ expansions give rise
to the same scenario  with the only difference being the value of $N_c(M)$
 which  varies with $M$.  In the physically important case of $M=N=3$,
 the $\epsilon$
%the principal chiral model, the $\epsilon$
 expansion predicts the absence of a stable chiral fixed point. 
Several Monte Carlo simulations~\cite{Lo-00,kz-93,Re-92,Ma-93} based essentially on Stiefel's models and ERG results~\cite{tmd-99} agree indicating a first-order phase transition for this system. Anyway
no data in the fixed dimension expansion exist.

In this paper we analyze the Hamiltonian (\ref{LGWH})
for $M \geq 3$ through a six-loop perturbative study in the framework of 
the fixed
dimension approach, in order to investigate the critical behavior of
frustrated spin systems
with nonplanar orderings directly in three dimensions and to have a
global picture of the region where the results obtained with different
perturbative methods are in agreement.

The paper is organized as follows. In Sec. \ref{sec2} we introduce the field-theoretical approach for the effective LGW Hamiltonian (\ref{LGWH}), in Sec. \ref{sec3} we present
the analysis method and we discuss the singularities of the Borel transform, 
 in Sec. \ref{sec4} we present the results and finally in Sec. \ref{sec5}
we draw our conclusions.

%%%%%%%%%%%%%%%%%%%%%%%%%%%%%%%%%%%%%%%%%%%%%%%%%%%%%%%%%%%%%%%%%%%%%%%%%%%
\section{Fixed dimension perturbative expansion of the O$(N)\times
 $O$(M)$ LGW Hamiltonian}
\label{sec2}

The Hamiltonian (\ref{LGWH}) is studied in the framework of the fixed dimension
 field-theoretical approach. As in the case of $XY$ and Heisenberg frustrated
models~\cite{Pelissetto:2000ne},
the idea is to perform an expansion in terms of the two quartic coupling
 constants $u_0$ and $v_0$, and to renormalize the theory imposing the
 following
 conditions on the (one-particle irreducible) two-point, four-point and
 two-point with an insertion of $\frac{1}{2} \phi^2$ correlation functions:

\begin{equation}
\Gamma^{(2)}_{ai,bj}(p) = 
\delta_{a,b}\delta_{i,j} Z_\phi^{-1} \left[ m^2+p^2+O(p^4)\right],
\label{ren1}  
\end{equation}
\begin{eqnarray}
\Gamma^{(4)}_{ai,bj,ck,dl}(0) = Z_\phi^{-2} m \left[  
{u\over 3}S_{ai,bj,ck,dl}
+ {v\over 6} \, C_{ai,bj,ck,dl}\right],
\label{ren2}
\end{eqnarray}
\begin{equation}
\Gamma^{(1,2)}_{ai,bj}(0) = \delta_{ai,bj} Z_t^{-1},
\label{ren3} 
\end{equation}
where $S_{ai,bj,ck,dl}$ and $C_{ai,bj,ck,dl}$ are appropriate combinatorial 
factors. \cite{Calabrese:2001be,Calabrese:2002bj}

The perturbative knowledge of the functions $\Gamma^{(2)}$, $\Gamma^{(4)}$ 
and $\Gamma^{(1,2)}$ allows one to relate  the renormalized parameters
 ($u,v,m$) to the bare ones ($u_0,v_0,r$ ).

The fixed points of the model are defined by the common zeros of the $\beta$
functions,
\begin{eqnarray}
\beta_u(u,v) = m \left. {\partial u\over \partial m}\right|_{u_0,v_0} ,
\quad \beta_v(u,v) = m \left. {\partial v\over \partial m}\right|_{u_0,v_0}
\end{eqnarray}
 and the stability properties of these points are determined by the eigenvalues 
$\omega_i$ of the matrix:
\begin{eqnarray}
\label{omega}
\Omega = \left(\begin{array}{ccc}
\displaystyle \frac{\partial \beta_u(u,v)}{\partial u} &
&\displaystyle \frac{\partial \beta_u(u,v)}{\partial v}\\
 \cr \displaystyle \frac{\partial \beta_v(u,v)}{\partial u}&
& \displaystyle  \frac{\partial \beta_v(u,v)}{\partial v} \end{array}
 \right).
\end{eqnarray}
A fixed point is stable if both the eigenvalues have a positive real part,
while a fixed point which possess eigenvalues with nonvanishing imaginary 
parts is called a focus.

The values of the critical exponents $\eta$, $\nu$  are
related to the RG functions $\eta_{\phi}$ and $\eta_t$ evaluated at 
the stable fixed point:

\begin{eqnarray}
\eta = \eta_\phi(u^*,v^*),
\label{eta_fromtheseries} 
\end{eqnarray}
\begin{eqnarray}
\nu = \left[ 2 - \eta_\phi(u^*,v^*) + \eta_t(u^*,v^*)\right] ^{-1},
\label{nu_fromtheseries}
\end{eqnarray}
where 
\begin{eqnarray}
\eta_\phi(u,v) &=& \left. {\partial \ln Z_\phi \over \partial \ln m}
       \right|_{u_0,v_0}
= \beta_u {\partial \ln Z_\phi \over \partial u} + 
\beta_v {\partial \ln Z_\phi \over \partial v} ,
\end{eqnarray}
\begin{eqnarray}
\eta_t(u,v) &=& \left. {\partial \ln Z_t \over \partial \ln m}
         \right|_{u_0,v_0}
= \beta_u {\partial \ln Z_t \over \partial u} + 
\beta_v {\partial \ln Z_t \over \partial v}.
\end{eqnarray}

The other critical exponents can be derived by scaling relations from $\eta$
 and $\nu$.

In this work the following symmetric rescaling of the coupling constants is adopted in order to obtain finite fixed point values in the limit of infinite components of the order parameter:

\begin{equation} 
u \equiv {16 \pi \over 3} R_{MN} \bar{u}, \qquad  v \equiv {16 \pi \over 3} R_{MN} \bar{v}
\end{equation}
where $R_N=9/(8+N)$.

For $N \rightarrow \infty$ four fixed points exist: Gauss $( \bar{u}=0,\bar{v}=0)$, Heisenberg $( \bar{u}=1,\bar{v}=0)$,
chiral $(\bar{u}=M,\bar{v}=M)$ which is stable, and antichiral $(\bar{u}=M-1,\bar{v}=M)$.
In the following we denote with $\bar{\beta}_{\bar{u}}$ and $\bar{\beta}_{\bar{v}}$ the $\beta$ functions corresponding to the above rescaled couplings.
%%%%%%%%%%%%%%%%%%%%%%%%%%%%%%%%%%%%%%%%%%%%%%%%
\section{Resummation and analysis method}
\label{sec3}

The perturbative six-loop series~\cite{vp} for the RG functions are 
 asymptotic and some resummation procedure is necessary  
in order to extract quantitative physical informations. In this work the
 property of Borel summability is exploited resumming the perturbative series
 by a Borel transformation combined with a conformal mapping~\cite{L-Z-77}
 (CM) which maps the domain of analyticity of the Borel transform 
(cut at the instanton singularity) onto a circle (see Refs.~\onlinecite{L-Z-77,Z-96}
for details).
This resummation procedure needs the knowledge of the large order behavior of
 the perturbative series which is connected with the singularity ($\bar{u}_b$)
closest to the origin at fixed $z=\bar{v}/ \bar{u}$.~\cite{Pelissetto:2000ne,Calabrese:2001be}
For the Hamiltonian (\ref{LGWH}) with generic O$(N) \times $O$(M)$ symmetry and
 $M\leq N$ we find

 \begin{eqnarray}
{1\over \ub_b} &=& - a\, R_{MN}
\quad  {\rm for} \quad  0< z<{2M \over M-1} ,
\label{bsing} \\
{1\over \ub_b} &=& - a\, R_{MN}\left[ 1- \left(1  - \displaystyle
{1\over M} \right)z \right]\,{\rm for} \,
 z<0 ,\, z>{2M \over M-1},
\nonumber
\end{eqnarray}
where $a=0.147777422$. 

In the case of $M=2$ the singularity reduces correctly to the one 
of Ref.~\onlinecite{Pelissetto:2000ne}.
For $z>{M \over M-1}$ there is a singularity on the real positive axis which
however is not the closest one to the origin for $z<{2M \over M-1}$. Thus, for
 $z>{M \over M-1}$ the series are not Borel summable. The region where the
 perturbative series are not Borel summable is the same of the one 
 where the condition for the boundedness of the free energy is satisfied. It is
 easy to show that the conditions are equivalent.

With this resummation procedure 
 one obtains a lot of approximants for each RG perturbative series
 characterized by three parameters $p$, $b$ and $\alpha$ which can be varied
in order to estimate the systematic errors in the final results. 
If $R(\ub,\vb)$ is a perturbative series in $\ub$ and $\vb$ 
\begin{equation}
R(\ub,\vb)= \, \sum_{k=0}^l \sum_{h=0}^{l-k} R_{hk} \ub^h \vb^k,
\end{equation}
the approximants will have the following form:
\begin{eqnarray}
E({R})_p(\alpha,b;\ub,\vb)= \sum_{k=0}^p 
  B_k(\alpha,b;\vb/ \ub) \nonumber\\
 \times   \int_0^\infty dt\,t^b e^{-t} 
  {y(\ub t;\vb/\ub)^k\over [1 - y(\ub t;\vb/ \ub)]^\alpha},
\label{approx}
\end{eqnarray}
where
\begin{equation} 
y(\ub;z) = {\sqrt{1 - \ub/\overline{u}_b(z)} - 1\over 
          \sqrt{1 - \ub/\overline{u}_b(z)} + 1}, 
\end{equation} 
and the coefficients $B_k$ are determined by the condition that the 
expansion of $E({R})_p(\alpha,b;\ub,\vb)$ in powers of $\ub$ and $\vb$ 
gives $R(\ub,\vb)$ to order $p$.

In order to find the fixed points of the Hamiltonian (\ref{LGWH}) the two
 $\bar{\beta}$ functions are resummed in the whole zone ($\bar{u}>0,\bar{v}>0$). For each $\bar{\beta}$ function the approximants are chosen which stabilize
 the series for small values of the coupling constants with varying the 
considered  perturbative order (number of loops), i. e. $\alpha=0,2,4$ and
 $b=3,6,9,12,15,18$.
The stability eigenvalues $\omega_i$ are evaluated by taking the 
 approximants for each $\bar{\beta}$ function and by computing the numerical 
derivatives of each couple of them at their common zero.
 Only the approximants which yield  fixed point coordinates compatible with
 their final values are considered.
 The critical exponents are computed by choosing the approximants which
 are more stable with varying the perturbative order (see Ref. \onlinecite{Calabrese:2002bj,Calabrese:2001be} for details).

\section{Results}
\label{sec4}
\begin{figure}[tb]
\includegraphics[width=8cm,height=6cm]{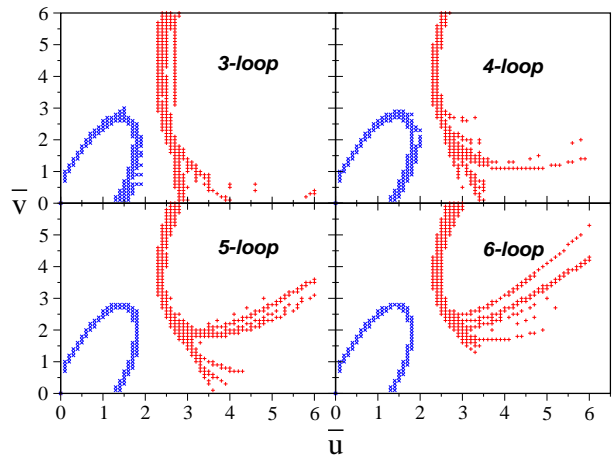}
\caption{Zeroes of the $\bar{\beta}$-functions for $N=3$ and $M=3$ in the
$(\bar{u},\bar{v})$ plane.
Pluses ($+$) and crosses ($\times$) correspond to zeroes of 
$\bar{\beta}_{\bar{v}}(\bar{u},\bar{v})$ and
$\bar{\beta}_{\bar{u}}(\bar{u},\bar{v})$ respectively.}
\label{m3n3}
\end{figure}

Fixing $M=3$, the analysis of the six-loop RG series reveals the absence of a
 stable chiral 
fixed point for the physical interesting case of $N=3$; in this case no 
second-order phase transition is expected as it is clear from the picture of
 the zeros of the $\bar{\beta}$ functions in Fig. \ref{m3n3}. In this Figure 
 the domain $0\leq \bar{u} \leq 6$,  $0\leq \bar{v} \leq 6$ is divided in $40^2$ rectangles, and all the sites in which at least two approximants for $\bar{\beta}_{\bar{u}}$ and $\bar{\beta}_{\bar{v}}$ vanish are marked.
These results are in agreement with the ERG prediction of 
Ref. \onlinecite{tmd-99} where no stable fixed 
point is found, and with Monte Carlo simulations on the corresponding Stiefel's
 model \cite{Lo-00,kz-93} which show a first-order phase transition.
Increasing the value of components of the order parameter $N$,
the curves of the zeros of the two $\bar{\beta}$ functions become closer and
 it exists a value $N_c(M=3)$ at which they intersect each other. For
$N>N_c(M=3)$ four fixed points exist in the region
$\bar{v}>0,\bar{u}>0$; the Gaussian one, the Heisenberg one, the antichiral
one and the stable chiral one. This last drives the system to a continuous 
phase transition.
For $N=N_c(M=3)$ the chiral and antichiral fixed point merge and for
$N<N_c(M=3)$  no stable fixed point is found.
The estimate of $N_c$ is obtained by  considering all the 324 possible 
combinations of the approximants for the two $\bar{\beta}$ functions and 
evaluating for each of them  the value of $N_c$. 
With this procedure we have $N_c(M=3)=11.1(6)$ at six-loop order.

The same scenario of fixed points holds for $M=4$ and $M=5$ with different 
values of $N_c$.

\begin{table}[b]
\caption{Comparison between values of $N_c(d=3,M)$ obtained with different 
perturbative methods.}
\begin{ruledtabular}
\begin{tabular}{c|c|c|c|c}
$M$ & 2 &3 & 4 & 5 \\
\hline
CM 5-loop &  &11.6(6) &15.4(9) &19.0(1.2)\\
CM 6-loop & 6.4(4)\cite{Calabrese:2003ib} &11.1(6) &14.7(8) &18(1)\\
$\epsilon$-expansion\cite{Pelissetto:2001fi} & 5.3(2) &9.1(9)&12(1)&14.7(1.6)\\
$1/N$-expansion\cite{Pelissetto:2001fi}& 5.3 & 7.3 & 9.2 &11.1\\
\end{tabular}
\end{ruledtabular}
\label{tabnc}
\end{table}
All the five-loop and six-loop estimates of $N_c$ for $M=3,\,4$ and $5$  are
 shown in Tab.~\ref{tabnc} where the $\epsilon$ expansion  and
 the $1/N$ predictions of Ref.~\onlinecite{Pelissetto:2001fi} are reported for
 a comparison.
 The agreement of the results is good even if it worsens with increasing 
 the magnitude of $M$. Nevertheless, in the case of a comparison
 with the $\epsilon$ expansion results, only three-loop series are 
 known which need to be resummed in order to extract some information. 
 Due to the low
perturbative order considered, one may only have an indication of the right 
results.  In fact the perturbative series can be evaluated extending
them analytically by means of a Pad\'e Borel transform~\cite{asv-95} (e. g. $N_c(M=2)\sim 3.39$) or by different techniques~\cite{Pelissetto:2001fi} (e. g. $N_c(M=2)\sim 5.3$) and  the results strongly depens on the method applied.
In the case of the $1/N$ expansion, one has to consider that $N_c$ is
 determined by perturbative series which are known only to $O(1/N^2)$, the 
correction to the leading term is not small, and since $N_c$ is expected not to
 be large in magnitude, one is extrapolating the results to value of $N$ which
 could be dangerous.
Furthermore the results of the different perturbative approaches are nearer in magnitude for small values of $M$ because with increasing the perturbative order
considered, the corrections to the leading terms become more important for
bigger $M$. Even the fixed-dimension results become less stable with increasing the value of $M$.
In order to provide a complete characterization of the critical properties 
of the Hamiltonian (\ref{LGWH}) we analyze it for 
large values of components of the order parameter at fixed $M$. The location 
of the chiral fixed point,  the stability eigenvalues and the critical exponents are calculated for $N=32, 64$ and are displayed in Tab.~\ref{tablargen}. 
For the evaluation of the critical exponents $\gamma$ and $\nu$ we resum 
the perturbative series of $1/ \gamma$ and $1/\nu$ because of their
better behavior. 
In these cases one can see a quite good agreement or the results with the $1/N$ expansion predictions.
The data of Tab.~\ref{tablargen} corroborate the expectation for which the line
 $N_c(M)$ limits the region where all the perturbative investigations bring to the same predictions.

\begin{table*}[t]
\caption{Large- $N$ results.}
\begin{ruledtabular}
\begin{tabular} {l|c|c|c|c}
$N=64,M=3$ & Ch. & $[\omega_1,\omega_2]$ & $\nu$ & $\gamma$ \\
\hline
C.M. (6-loop) & [2.934(1),3.0051(8)]   & [0.960(1),0.915(2)] &  0.960(2)&1.911(4) \\
 C.M. (5-loop)    & [2.935(2),3.006(2)] & [0.958(2),0.911(3)] & & \\
$1/N$-expansion~\cite{Gracey:2002ze,Pelissetto:2001fi} &    & [0.97,0.93]  & 0.965& 1.92\\
\tableline\hline
$N=32,M=3$ & Ch. & $\omega_1,\omega_2$ &  $\nu$ & $\gamma$ \\
\hline
  C.M. (6-loop)    & [2.861(2),3.012(2)] & [0.919(2),0.810(6)] & 0.912(4)&1.807(8) \\
C.M. (5-loop)    & [2.862(7),3.014(7)] & [0.916(5),0.80(1)] && \\
$1/N$-expansion~\cite{Gracey:2002ze,Pelissetto:2001fi} &  &[0.94,0.86]& 0.927&1.84 \\
\tableline\hline
$N=64,M=4$ & Ch. & $\omega_1,\omega_2$ &  $\nu$ & $\gamma$ \\
\hline
 C.M. (6-loop)    & [3.8550(15),3.950(1)] & [0.9536(11),0.891(2)] & 0.949(3)&1.886(6) \\
C.M. (5-loop)    & [3.856(4),3.951(3)] & [0.952(3),0.886(4)] &&  \\
$1/N$-expansion~\cite{Gracey:2002ze,Pelissetto:2001fi} &  &[0.97,0.91]& 0.956& 1.90\\
\tableline\hline
$N=32,M=4$ & Ch. & [$\omega_1,\omega_2$] & $\nu$& $\gamma$  \\
 \hline
 C.M. (6-loop) & [3.7035(30),3.905(3)]   & [0.905(2),0.75(1)] &  0.887(6) &1.75(1)\\
 C.M. (5-loop)    & [3.65(7),3.906(11)] & [0.904(6)0.73(2)] &&\\
$1/N$-expansion~\cite{Gracey:2002ze,Pelissetto:2001fi} &    & [0.93,0.83]  &0.907& 1.80\\
\tableline\hline
$N=64,M=5$ & Ch. & $\omega_1,\omega_2$ & $\nu$& $\gamma$  \\
\hline
  C.M. (6-loop)    & [4.7676(16),4.8864(15)] & [0.9467(12),0.866(3)] &0.937(3) &1.862(6)\\
 C.M. (5-loop)    & [4.770(6),4.888(5)] & [0.945(3),0.860(7)]  && \\
$1/N$-expansion~\cite{Gracey:2002ze,Pelissetto:2001fi} &  &[0.96,0.90]&0.946& 1.88 \\
\tableline\hline
$N=32,M=5$ & Ch. & [$\omega_1,\omega_2$] & $\nu$ & $\gamma$ \\
 \hline
 C.M. (6-loop) & [4.528(6),4.783(3)]   & [0.8917(20),0.675(15)]  &0.860(8)&1.696(16) \\
 C.M. (5-loop)    & [4.52(2),4.78(2)] & [0.893(6),0.65(2) &&\\
$1/N$-expansion~\cite{Gracey:2002ze,Pelissetto:2001fi} &    & [0.92,0.80]  &0.887&1.75\\
\end{tabular}
\label{tablargen}
\end{ruledtabular}
\end{table*}

\section{Conclusions}
\label{sec5}
In this work we have studied the critical thermodynamics of the
 O$(N)\times$O($M$) spin model for $N \geq M \geq 3$ and $M\leq 5$ on the
 basis of six-loop RG series  in fixed dimension. We have evaluated the large 
order 
behavior of the perturbative RG series and we have analyzed them by means of
a CM resummation technique. For the physically interesting case of 
%the principal chiral model
$N=M=3$, no stable fixed point exists in the 
RG flow diagram leading to the conclusion that this system is expected to
 undergo a first-order phase transition. This result agrees with previous ERG 
predictions~\cite{tmd-99} and Monte Carlo studies on Stiefel's 
models.~\cite{Lo-00,kz-93}
For $M=3,4,5$ we have computed the critical value $N_c(M)$ of components of 
the order parameter which separates the region of first-order phase transition 
to the second-order one, finding $N_c(M=3)=11.1(6)$, $N_c(M=4)=14.7(8)$ and
$N_c(M=5)=18(1)$. In this case the comparison with the $\epsilon$,~\cite{Pelissetto:2001fi}
 and $1/N$ expansion~\cite{Pelissetto:2001fi}
predictions is quite good even if it worsens with increasing the magnitude of
$M$. 
Finally, being the region $N>N_c(M)$ expected to be connected with that
 accessible by the
large $N$ analysis, we have given a characterization of the critical properties
 of these systems for large values of $N$ finding a satisfactory agreement 
between the two perturbative methods.

\section*{ACKNOWLEDGMENTS}
 The author is grateful to A. Pelissetto and E. Vicari for giving 
him the six-loop perturbative expressions of the RG functions, and to P.~Calabrese, P. Rossi, A. I. Sokolov and E.~Vicari for a critical reading of this manuscript and useful discussions.


\begin{references}


\bibitem{Kawamura-98}
H.~Kawamura, J. Phys.: Condens. Matter {\bf 10}, 4707 (1998).
%%CITATION = JCOME,10,4707;%%



%\cite{Pelissetto:2000ek}
\bibitem{Pelissetto:2000ek}
A.~Pelissetto and E.~Vicari,
%``Critical phenomena and renormalization-group theory,''
Phys.\ Rept.\  {\bf 368}, 549 (2002).
%[arXiv:cond-mat/0012164].
%%CITATION = COND-MAT 0012164;%%


\bibitem{cp-97}
M.~F. Collins and O.~A. Petrenko,
Can. J. Phys. {\bf 75}, 605 (1997).

\bibitem{Kawamura-88}
H.~Kawamura, Phys. Rev. B {\bf 38}, 4916 (1988);
erratum  B {\bf 42}, 2610 (1990).
%%CITATION = PHRVA,B38,4916;%%

\bibitem{Ka-90}
H.~Kawamura, J. Phys. Soc. Japan {\bf 59}, 2305 (1990);



\bibitem{asv-95}
S.~A.~Antonenko, A.~I.~Sokolov, and K.~B.~Varnashev,
Phys. Lett. A {\bf 208}, 161 (1995).


%\cite{Pelissetto:2001fi}
\bibitem{Pelissetto:2001fi}
A.~Pelissetto, P.~Rossi and E.~Vicari,
%``Large-n critical behavior of O(n) x O(m) spin models,''
Nucl.\ Phys.\ B {\bf 607}, 605 (2001).
%[arXiv:hep-th/0104024].
%%CITATION = HEP-TH 0104024;%%



%\cite{Gracey:2002ze}
\bibitem{Gracey:2002ze}
J.~A.~Gracey,
%``Critical exponent omega at O(1/n) in O(n) x O(m) spin models,''
Nucl.\ Phys.\ B {\bf 644}, 433 (2002).
%[arXiv:hep-th/0209053].
%%CITATION = HEP-TH 0209053;%%

%\cite{Gracey:2002pm}
\bibitem{Gracey:2002pm}
J.~A.~Gracey,
%``Chiral exponents in O(N) x O(m) spin models at O(1/N**2),''
Phys. Rev. B {\bf 66}, 134402 (2002).
%[arXiv:cond-mat/0208309].
%%CITATION = COND-MAT 0208309;%%


\bibitem{az-de-de-jo}
P.~Azaria, B.~Delamotte and T. Jolicoeur,
Phys. Rev. Lett. {\bf 64}, 3175 (1990);
P.~Azaria, B.~Delamotte, F.~Delduc, and Th.~Jolicoeur,
 Nucl. Phys. B {\bf 408}, 585 (1993).


%\cite{Calabrese:2001be}
\bibitem{Calabrese:2001be}
P.~Calabrese and P.~Parruccini,
%``The critical behavior of 2-d frustrated spin models with noncollinear order,''
Phys.\ Rev.\ B {\bf 64}, 184408  (2001).
%[arXiv:cond-mat/0105551].
%%CITATION = COND-MAT 0105551;%%


%\cite{Calabrese:2002bj}
\bibitem{Calabrese:2002bj}
P.~Calabrese, E.~V.~Orlov, P.~Parruccini and A.~I.~Sokolov,
%``Chiral critical behavior in two dimensions from five-loop renormalization-group expansions,''
Phys. Rev. B {\bf 67}, 024413 (2003).
%[arXiv:cond-mat/0207187].
%%CITATION = COND-MAT 0207187;%%


%\cite{Calabrese:2002af}
\bibitem{Calabrese:2002af}
P.~Calabrese, P.~Parruccini and A.~I.~Sokolov,
%``Chiral phase transitions: focus driven critical behavior in systems with planar and vector ordering,''
Phys.\ Rev.\ B {\bf 66}, 180403 (2002).
%[arXiv:cond-mat/0205046].
%%CITATION = COND-MAT 0205046;%%



%\cite{Calabrese:2003ib}
\bibitem{Calabrese:2003ib}
P.~Calabrese, P.~Parruccini and A.~I.~Sokolov,
%``Critical thermodynamics of three-dimensional chiral model for N > 3,''
Phys.\ Rev.\ B {\bf 68}, 094415 (2003).
%%CITATION = COND-MAT 0304154;%%






%\cite{Pelissetto:2000ne}
\bibitem{Pelissetto:2000ne}
A.~Pelissetto, P.~Rossi and E.~Vicari,
%``The critical behavior of frustrated spin models with noncollinear order,''
Phys.\ Rev.\ B {\bf 63}, 140414 (2001).
%[arXiv:cond-mat/0007389].
%%CITATION = COND-MAT 0007389;%%




%\cite{Pelissetto:2001hm}
\bibitem{Pelissetto:2001hm}
A.~Pelissetto, P.~Rossi and E.~Vicari,
%``Chiral exponents in frustrated spin models with noncollinear ordering,''
Phys.\ Rev.\ B {\bf 65}, 020403 (2002).
%[arXiv:cond-mat/0106525].
%%CITATION = COND-MAT 0106525;%%


\bibitem{as-94}
S.~A.~Antonenko and A.~I.~Sokolov,
Phys. Rev. B {\bf 49}, 15901 (1994);
D.~Loison, A.~I.~Sokolov, B.~Delamotte, S.~A.~Antonenko, K.~D.~Schotte, and
H.~T.~Diep,
JETP Lett. {\bf 72}, 337 (2000).


%\cite{Tissier:2001uk}
\bibitem{Tissier:2001uk}
M.~Tissier, B.~Delamotte and D.~Mouhanna,
%``XY frustrated systems: continuous exponents in discontinuous phase transitions,''
Phys. Rev. B {\bf 67}, 134422 (2003);
%``Heisenberg frustrated magnets: a nonperturbative approach,''
Phys. Rev. Lett. {\bf84}, 5208 (2000);
%``An exact renormalization group approach to frustrated magnets,''
Int.\ J.\ Mod.\ Phys.\ A {\bf 16} (2001) 2131.





\bibitem{tmd-99}
M.~Tissier, D.~Mouhanna, B.~Delamotte,
Phys. Rev. B {\bf 61}, 15327 (2000).
%[arXiv:cond-mat/9908352].






\bibitem{zu-94}
G.~Zumbach, 
Nucl. Phys. B {\bf 413}, 771 (1994).








\bibitem{Re-92}
J.~N.~Reimers, J.~E.~Greedan, and M.~Bj\"orgvinsson, 
Phys. Rev. B {\bf 45}, 7295 (1992).

\bibitem{Ma-93}
A.~Mailhot and M.~L.~Plumer,Phys. Rev. B {\bf 48}, 9881 (1993).



\bibitem{Lo-00}
D.~Loison, 
Eur. Phys. J. B {\bf 15}, 517 (2000).

\bibitem{kz-93}
H.~Kunz, G.~Zumbach,
J.~Phys. A {\bf 26}, 3121 (1993).










%\cite{vp}
\bibitem{vp}
The six-loop perturbative series for the RG functions were computed by
E.~Vicari and A.~Pelissetto in an unpublished work and were given to the author
in a Private Communication. 
The series are available on request to the author.


\bibitem{L-Z-77} J.~C.~Le Guillou and J.~Zinn-Justin,
Phys.\ Rev.\ Lett.\ {\bf 39}, 95 (1977);
Phys.\ Rev.\ B {\bf 21}, 3976 (1980).
%%CITATION = PRLTA,39,95;%%
%%CITATION = PHRVA,B21,3976;%%


\bibitem{Z-96}  J.~Zinn-Justin, {\em Quantum Field Theory and 
Critical Phenomena}, 3rd edition (Clarendon Press, Oxford, 1996).















\end{references}
\end{document}